\documentclass[aps,showpacs,preprint]{revtex4}%
\usepackage{graphicx} 
\usepackage{dcolumn}  
\usepackage{bm}       
\usepackage{epsfig}
\usepackage{comment}
\usepackage{color}
\usepackage{multirow}
\usepackage{pgfrcs}

\begin{document}

\begin{titlepage}
\title{Electronic and magnetic properties of FeSe$_{0.5}$Te$_{0.5}$: A first-principles study}
\author{Menglei Li,$^{1,2}$ Fawei Zheng,$^1$ and Ping Zhang$^1$$^,$\footnote{zhang\underline{ }ping@iapcm.ac.cn}}
\affiliation{$^1$Institute of Applied Physics and Computational Mathematics, P.O. Box 8009, Beijing 100088, P.R. China\\
$^2$Center for Fusion Energy Science and Technology, Chinese Academy of Engineering Physics, Beijing 100084, P.R. China}

\date{\today}

\begin{abstract}
The atomic structures, electronic band structures and magnetic properties of monolayer FeSe and FeSe$_{0.5}$Te$_{0.5}$ of different configurations have been systematically investigated via first-principles calculations with the inclusion of spin-orbit coupling (SOC). Three different antiferromagnetic (AFM) orders, including checkerboard order, collinear order and pair-checkerboard order, as well as paramagnetic state have been explored. In monolayer FeSe, collinear AFM order is found to be the most stable order, in accordance with previous investigations. Substituting half Se atoms with Te atoms, the pair-checkerboard AFM order is the ground-state magnetic order in FeSe$_{0.5}$Te$_{0.5}$. Both AFM-ordered FeSe and FeSe$_{0.5}$Te$_{0.5}$ have Dirac-cone-like band structures. SOC has a great influence on the band structures at the Dirac cone. The direction of the magnetic moments (in-plane or out-of-plane) directly determines whether the Dirac cone could be opened by SOC, and the gap values also relate to the specific magnetic structure. Although SOC is stronger in FeSe$_{0.5}$Te$_{0.5}$, the SOC-induced band gaps are either only slightly enlarged or even much shrunk compared with those gaps in FeSe. Due to the symmetry breaking brought by Te-substitution, the band structures of FeSe$_{0.5}$Te$_{0.5}$ have a new feature of combined Rashba-Dresselhaus splitting.
Our results have provided a comprehensive study on the magnetic property of FeSe$_{0.5}$Te$_{0.5}$, which may help to understand the relation between magnetism and the superconductivity in the high-$T_{C}$ monolayer superconductor.


\end{abstract}

\maketitle

\draft

\vspace{2mm}

\end{titlepage}

\section{Introduction}

Recently, monolayer FeSe grown on SrTiO$_3$ substrate has attracted a lot of attention for its high-$T_C$ superconductivity, which has not been found in bulk FeSe in spite of similar structures \cite{monolayer1,monolayer2,monolayer3}. Some researches demonstrate that the different superconductive behaviours of monolayer FeSe and bulk FeSe are due to the influence of the SrTiO$_3$ substrate and the interface between FeSe and SrTiO$_3$ \cite{berlijn2014,Cao2014,Cui2015,interfaceARPES}, while the mysterious antiferromagnetic (AFM) structures of FeSe also paly a major role.
Therefore, theoretical studies on the magnetic property of FeSe have been intensively carried out, and a variety of AFM orders, including checkerboard (CB) AFM \cite{zheng}, collinear (COL) AFM \cite{xiangtao} and pair-checkerboard (PCB) AFM orders \cite{Cao2015}, have been investigated. A novel nematic AFM phase is also proposed theoretically \cite{Lu}. However, the magnetic measurements such as neutron scattering \cite{neutron} and superconducting quantum interference device (SQUID) \cite{squid}, can only detect the magnetic signal of bulk FeSe rather than the FeSe monolayer. In monolayer FeSe, only the electronic structure can be obtained via, for example, the angle resolved photoemission spectroscopy (ARPES) and scanning tunnelling spectroscopy \cite{ARPES2,interfaceARPES,FengLiu}. The comparison between experimental and theoretical achievements are confusing. On one hand, the band structures of CB-AFM and PCB-AFM ordered FeSe resemble the results from the ARPES experiments \cite{cohen,zheng,Cao2015,zheng2}. On the other hand, COL-AFM order is the ground-state AFM order from first-principles calculations \cite{xiangtao}. These controversial results suggest that the magnetic structure of monolayer FeSe is intriguing, and still needs to be further confirmed.

In bulk FeSe, the Te substitution for Se atoms can enhance the critical temperature from 9 K \cite{bulkTC} to more than 14 K \cite{FeSeTeTC,FeSeTeTC2}. Therefore, it is natural to expect that the monolayer FeSe$_{1-x}$Te$_x$ also has high-$T_C$ superconductivity. Indeed, some experiments have demonstrated the high-$T_C$ superconductive behaviour in FeSe$_{1-x}$Te$_x$ and the critical temperature is found to be dominated by grown conditions and in-plane strain \cite{external}. Moreover, FeSe$_{1-x}$Te$_x$, as well as FeSe, is a topological insulator \cite{topoFeSeTe}. The topological character combined with AFM structures and superconductivity could give rise to a rich variety of novel physical phenomena. But a systematical analysis of the magnetic properties and band structures of FeSe$_{1-x}$Te$_x$ in theory is still lacked. Bearing this in mind, we aim at providing a comprehensive study on the electronic property of FeSe$_{1-x}$Te$_{x}$ with different magnetic orders.

In this paper, we calculate the electronic band structures of monolayer FeSe and typical Te-substituted monolayer FeSe$_{0.5}$Te$_{0.5}$ with different magnetic orders using first-principles tools, and search the magnetic ground state for both monolayer FeSe and FeSe$_{0.5}$Te$_{0.5}$. 
The direction of the magnetic moments can make a big difference on the band structures due to the spin-orbit interaction. Besides, since Te-substitution breaks the space symmetry to some extent, the band structures of FeSe$_{0.5}$Te$_{0.5}$ have a novel Rashba-Dresselhaus splitting. But the band gaps are either only slightly enlarged (with PCB order) or even much shrunk (with COL order) in FeSe$_{0.5}$Te$_{0.5}$ compared with those in FeSe.

This paper is organized as follows. First we describe the calculational details and geometries of different structures. Then we present our results for pristine monolayer FeSe, followed by our main results in FeSe$_{0.5}$Te$_{0.5}$. At last, there is a summary.

\begin{figure}[ptb]
\begin{center}
\includegraphics[width=0.58\linewidth]{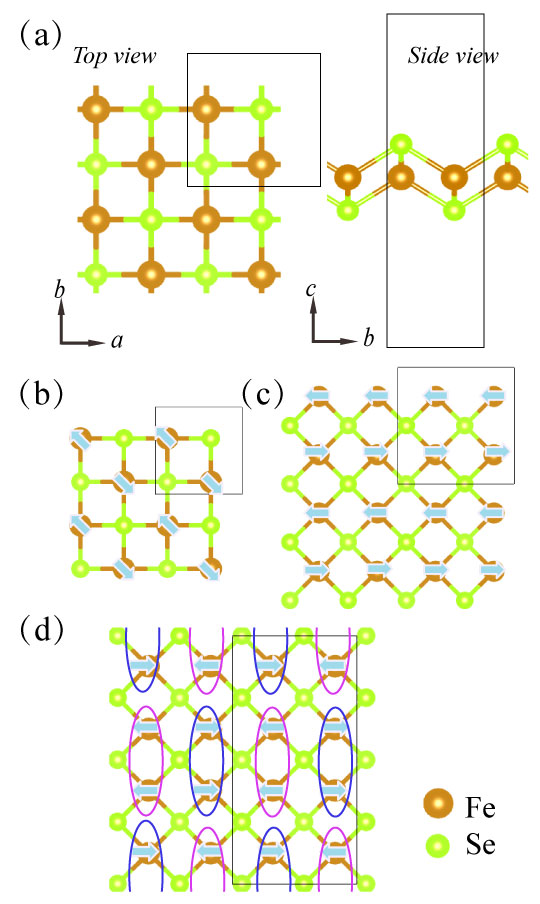}
\end{center}
\caption{(Color online) (a) The structure of a FeSe monolayer in top view and side view. Three calculated AFM orders: (b) checkerboard AFM, (c) collinear AFM and (d) pair-checkerboard AFM. The light-blue arrows represent the magnetic moment of the Fe atom and the black lines show the boundaries of the magnetic unit cells in our calculations.}
\label{Fig1}%
\end{figure}

\section{Methodology}

We have performed spin-polarized first-principles calculations on the total energy, magnetic and electronic structures of FeSe and FeSe$_{0.5}$Te$_{0.5}$ with the project augmented wave method \cite{paw1,paw2} as implemented in the VASP code \cite{vasp}. Plane waves less than 400 eV were used to expand the wave functions. For the exchange-correlation potential, we have adopted the generalized gradient approximation (GGA) of Perdew-Burke-Ernzerhof (PBE) type \cite{pbe1}. The spin-orbit coupling (SOC) effect was included all through the calculations; meanwhile we have also performed normal calculations without SOC for comparison. Since SOC distinguishes the direction of the electron spins, the magnetic moments can thus rotate three-dimensionally. In view of this, both situations in which the magnetic moments are parallel to the monolayer (in-plane) and perpendicular to the monolayer (out-of-plane) were considered. The structural optimization has employed the conjugate gradient algorithm \cite{cg} and the residual forces on all relaxed atoms were smaller than 0.01 eV/\AA. The monolayer FeSe and FeSe$_{0.5}$Te$_{0.5}$ were both stacked with a vacuum layer of 20 \AA\ thick to eliminate the coupling between periodic monolayers. During the structural relaxation, the lattice parameters were fixed at $a = b =$ 3.9015 \AA\ to mimic the confinements of SrTiO$_3$ substrate. In the simplest AFM ordered FeSe, i.e. FeSe with the CB order, the magnetic unit cell is $\sqrt{2}\times\sqrt{2}$ of the original non-magnetic unit cell and contains 4 atoms. For FeSe with COL and PCB orders, the magnetic unit cells are $2\times 2\times 1$ of the original unit cell containing 8 atoms and $2\times 4\times 1$ enlarged cell containing 16 atoms, respectively, as shown in Fig. 1. Accordingly, the k-grids for CB-, COL- and PCB-AFM phases are $9\times 9\times 1$, $7\times 7\times 1$ and $9\times 5\times 1$ in a Monkhorst-Pack grids, respectively. Besides, a paramagnetic (PM) monolayer is also investigated with the same unit cell as CB ordered monolayer for reference.

\begin{figure}[!ptb]
\begin{center}
\includegraphics[width=0.7\linewidth]{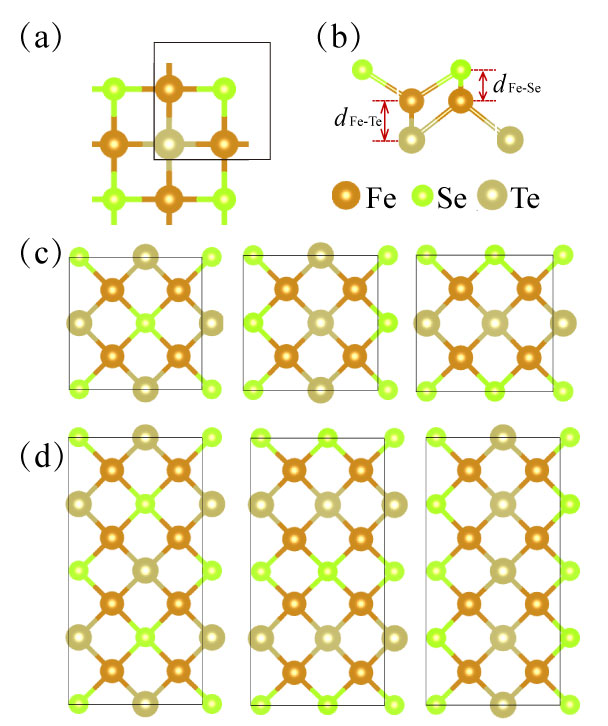}
\end{center}
\caption{(Color online) Different FeSe$_{0.5}$Te$_{0.5}$ configurations in (a) paramagnetic (PM) state and CB-AFM order, (c) COL-AFM order and (d) PCB-AFM order. The distance between the Se (Te) atom and the Fe-plane is denoted as $d_{\mathrm{Fe-Se}}$ ($d_{\mathrm{Fe-Se}}$) as shown in (b). The Te-substituted configurations in the left panel, middle panel and right panel of (c), are denoted as Struc.1, Struc.2 and Struc.3 for the COL ordered system. The same goes to (d).}
\label{Fig2}%
\end{figure}

For Te substitution, considering that Fe ions with antiparallel magnetic moments are not equivalent, there could be more than one configurations as shown in Fig.~2. Paramagnetic FeSe$_{0.5}$Te$_{0.5}$ and CB-AFM FeSe$_{0.5}$Te$_{0.5}$ each have only one configuration, in which all Se atoms are on top of the Fe-layer and all Te atoms are below the Fe-layer, or the other way around, as shown in Fig.~2(a). For FeSe$_{0.5}$Te$_{0.5}$ with COL and PCB orders, besides the configuration mentioned above as shown in the leftmost panel of Fig.~2(c,d), Te atoms can both take the top-layer and bottom-layer positions simultaneously. Since Fe ions with magnetic moments in different directions are not equivalent, the substituted Te atoms can align along $a$-axis or $b$-axis. Therefore, with COL order or PCB order, there are 3 types of Te-substituted FeSe$_{0.5}$Te$_{0.5}$ configurations which for the later convenience are denoted as Struc.1, Struc.2 and Struc.3.

\section{Results}
\subsection{Properties of Pure FeSe}

For the monolayer FeSe, the lattice constant is fixed at the calculated lattice parameter of SrTiO$_3$, which is $a = b =$ 3.9015 \AA. The total energies for monolayer FeSe with different magnetic orders are listed in Tab.\ref{tab:table 1}, which clearly shows that FeSe with COL-AFM order is more stable than other calculated magnetic states, in accordance with previous first-principles calculations. Including the SOC effect, the total energies for all the magnetic states are lowered by around 60 meV and FeSe with COL-AFM order still being the most stable one. In the AFM ordered monolayer FeSe, the energy difference between FeSe with in-plane magnetization and out-plane magnetization is no more than 1 meV, as presented in the last two columns of Tab.\ref{tab:table 1}, revealing that the magnetic anisotropy energy should be quite small. However, the magnetization still tends to be parallel to the Fe-plane rather than perpendicular to it, by a tiny energy difference varying from 0.4 meV to 1 meV depending on the particular AFM order.

\begin{table}[!bp]
\centering
\caption{\label{tab:table 1} The total energies of monolayer FeSe with different magnetic orders. (The energies are in units of eV averaged on per FeSe formula cell)}
\begin{ruledtabular}
\begin{tabular}{c|ccc}
Magnetic & No SOC &\multicolumn{2}{c}{Including SOC } \\
Order &  & In-plane & Out-of-plane \\
\hline
PM & -24.5082 & -24.5601 & -24.5601  \\
CB AFM & -24.7089 & -24.7611 & -24.7601  \\
COL AFM & -24.8282 & -24.8798 & -24.8791  \\
PCB AFM & -24.8156 & -24.8669 & -24.8665  \\
\end{tabular}
\end{ruledtabular}
\end{table}

Next, the band structures of monolayer FeSe with different magnetic orders are demonstrated in Fig.~3. Note that the unit cells of different magnetic states are not the same, so the first Brillouin zones and k points are different. For instance, the M point of FeSe unit cell in PCB-AFM state is the middle point on the X'-M path of the unit cell with COL-AFM order.

\begin{figure*}[!tbp]
\includegraphics[width=1.0\textwidth]{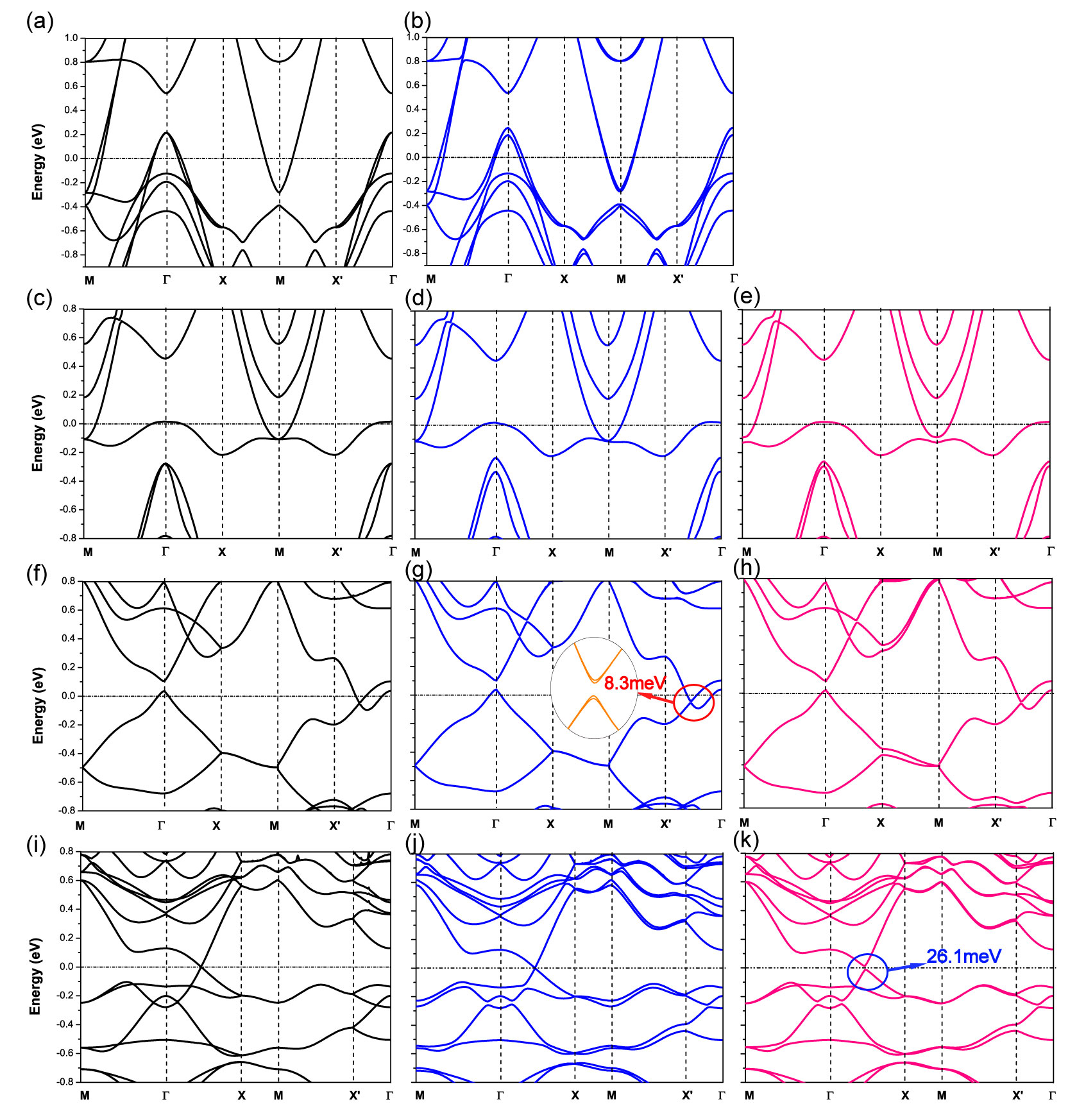}
\caption{(Color online) Band structures for monolayer FeSe with (a-b) paramagnetic order, (c-e) checkerboard AFM order, (f-h) collinear AFM order and (i-k) pair-checkerboard AFM order. The band structures in black lines [(a), (c), (f), (i)] are calculated not including spin-orbit coupling, and the band structures from two spin channels are exactly the same. Those bands plot in blue and pink lines are calculated including SOC effect with in-plane [(d), (g), (j)] and out-of-plane [(e), (h), (k)] magnetizations respectively. For paramagnetic FeSe, the direction of magnetization does not make any difference on the band structure at all, as in (b).}
\label{Fig3}%
\end{figure*}

For the paramagnetic monolayer FeSe, there is an electron-like Fermi pocket at M point. Apparently, SOC destroys the degeneracy of the bands at $\Gamma$ and M to a certain extent. As is well known, the magnetic order plays an important role in the character of the band structures. In monolayer FeSe with CB-AFM order, the band structure changes a lot with a flat band crossing the Fermi level. The degeneracy is also removed by SOC at $\Gamma$ as in paramagnetic FeSe. But now the direction of the magnetization alters the properties around M point. The normal calculation and calculation with SOC effect and in-plane magnetization result in the contact between the flat band and the electron pocket at M point, while SOC with the out-of-plane magnetic moment opens a band gap of 12.8 meV at M, as shown in Fig.~3(c-e). This result is in agreement with Ref.\onlinecite{FengLiu}, according to which the on-site energy of $d_{z^2}$ orbital is different with that of $d_{xz}$/$d_{yz}$ orbital due to the crystal field effect and the gap at the M point is opened by an on-site coupling between the $d_{xz}$ and $d_{yz}$ orbitals of the Fe atoms. Although the band structures can be such different, the magnetic anisotropy is not significant since the energy difference of CB-AFM states with in-plane and out-of-plane magnetization is only 1 meV as mentioned in the last section.

As to the band structures of the COL-AFM state, a notable feature is the Dirac cone emerging in the midst of the $\Delta$ line. However, unlike the situations of PM and CB-AFM states, SOC only changes the band structures limitedly. Comparing Fig.~3(f) with Fig.~3(g, h), there are two major distinct places. The degeneracy along the Y line is lifted by out-of-plane magnetization as shown Fig.~3(h), while the in-plane magnetization opens a gap of 8.4 meV at the Dirac cone as manifested in Fig.~3(g).

At last, the band structures of PCB-AFM are shown in Fig.~3(i-k). Our results are highly accordant with those from Cao et.al \cite{Cao2015}, yielding an exotic Dirac-cone-like feature. Under the effect of SOC with out-of-plane magnetization, monolayer FeSe becomes insulating with a band gap of 26.1 meV. This phenomenon is explained in a way that $d_{xz}$ and $d_{yz}$ orbitals which originally belong to different symmetry groups can now be mixed by SOC \cite{Cao2015}. However, when the magnetization is parallel to the Fe-plane, there is no such gap-opening, just like the case of CB-AFM state. In the light of the explanation for gap-opening mechanism in the system with out-of-plane magnetization, what happened with in-plane magnetization can be explained. The difference between $d_{xz}$ and $d_{yz}$ orbitals are maintained; in other words, $d_{xz}$ and $d_{yz}$ still belong to different symmetry groups in the existence of the in-plane magnetic moment. Therefore, SOC cannot open a gap and the band structure is still metallic possessing the Dirac-cone-like feature. In contrast, the in-plane magnetization, rather than the out-of-plane magnetization, can lift the degeneracy along X-M-X' direction, indicating that the in-plane magnetization can give rise to other kind of symmetry breaking.

\subsection{Properties of FeSe$_{0.5}$Te$_{0.5}$}
\subsubsection{Total Energy}

Before we turn to the electronic structures of monolayer FeSe$_{0.5}$Te$_{0.5}$, we will first discuss the calculated total energies for FeSe$_{0.5}$Te$_{0.5}$ with different magnetic orders as in pure FeSe. Those energies are listed in Tab.\ref{tab:table 2}. Note that for COL order and PCB order, there are three substitutional structures and thus Tab.\ref{tab:table 2} has more rows than Tab.\ref{tab:table 1}. The spin-orbit interaction is doubled in FeSe$_{0.5}$Te$_{0.5}$ than in FeSe. When SOC is included, total energy can be lowered by about 120 meV in FeSe$_{0.5}$Te$_{0.5}$, while the energy lowering is only 60 meV in pure FeSe. This is easy to understand because a Te atom is much heavier than a Se atom, which necessarily leads to a stronger spin-orbit interaction. However, the magnetic anisotropy energy has not changed from that of pristine FeSe, that is, the energy difference between structures with in-plane magnetization and out-of-plane magnetization is still as small as 0.1 to 0.2 meV.

\begin{table}[!bp]
\centering
\caption{\label{tab:table 2} The total energies of monolayer FeSe$_{0.5}$Te$_{0.5}$ of different configurations with different magnetic orders from calculations with and without SOC effect. The energies are in units of eV averaged for per FeSe$_{0.5}$Te$_{0.5}$ formula cell.}
\begin{ruledtabular}
\begin{tabular}{cc|ccc}
\multicolumn{2}{c|}{Magnetic} & No SOC &\multicolumn{2}{c}{Including SOC } \\
\multicolumn{2}{c|}{Order} &  & In-plane & Out-of-plane \\
\hline
\multicolumn{2}{c|}{PM} & -23.6945 & -23.8237 & -23.8237  \\ \hline
\multicolumn{2}{c|}{CB-AFM} & -23.8973 & -24.0284 & -24.0279  \\ \hline
\multirow{3}{*}{COL-AFM }& Struc. 1 & -23.9846 & -24.1155 & -24.1139  \\
& Struc. 2 & -23.9890 & -24.1203 & -24.1190  \\
& Struc. 3 & -23.9881 & -24.1198 & -24.1182  \\ \hline
\multirow{3}{*}{PCB-AFM }& Struc. 1 & -23.9977 & -24.1280 & -24.1279  \\
& Struc. 2 & -24.0218 & -24.1522 & -24.1521  \\
& Struc. 3 & -24.0024 & -24.1331 & -24.1332  \\
\end{tabular}
\end{ruledtabular}
\end{table}

For COL-AFM and PCB-AFM FeSe$_{0.5}$Te$_{0.5}$, three kinds of substitutional configurations are not equivalent in energy. The first type of configuration, Struc.1, in which all the Se atoms are on top of the Fe-plane while all the Te atoms are on the bottom, is the most unstable configuration for both COL state and PCB state. But the energies of Struc.2 and Struc.3 with COL-AFM order are quite close with a difference of less than 1 meV. Yet this is not the case in PCB-AFM state, where the energy difference between Struc.2 and Struc.3 is much larger than that between Struc.1 and Struc.3. In other words, the Struc.2 is far more stable than the other two substitutional structures in PCB-AFM ordered FeSe$_{0.5}$Te$_{0.5}$. This discrepancy is probably due to the different symmetries of the magnetic unit cell of COL-AFM and PCB-AFM states.

\subsubsection{Structure and Bond length}

The change of atomic structures after Te-substitution is explored in terms of the average distance between Te/Se atoms and Fe-plane (i.e. $d_{\mathrm{Fe-Se}}$ and $d_{\mathrm{Fe-Te}}$), as shown in Tab. \ref{tab:table 3}. The influence of SOC on the atomic structure is quite tiny, as including SOC the relaxed $d_{\mathrm{Fe-Se}}$/$d_{\mathrm{Fe-Te}}$ changes by less than 0.002 \AA. Therefore, we only need to focus on $d_{\mathrm{Fe-Se}}$ and $d_{\mathrm{Fe-Te}}$ resulted from calculations without SOC.

As shown in Tab.\ref{tab:table 3}, no matter in pure FeSe or in FeSe$_{0.5}$Te$_{0.5}$, the paramagnetic state has the shortest $d_{\mathrm{Fe-Se}}$. With different AFM orders, the distance becomes further compared to the PM state owing to the magnetic interaction. Among the three AFM orders, the length of $d_{\mathrm{Fe-Se}}$ is the least in CB-AFM order and the largest in PCB-AFM order. Generally, Te-substitution could shorten $d_{\mathrm{Fe-Se}}$. But in Struc.2 and Struc.3 with COL-AFM order or PCB-AFM order, which happen to be two more stable substitutional structures with the certain magnetic order, $d_{\mathrm{Fe-Se}}$ slightly increases.
As to the distance between Te atoms and Fe-plane, it is natural that Te atoms tend to move farther from Fe-plane than Se atoms since Te atom has a larger ionic radius. Therefore, $d_{\mathrm{Fe-Te}}$ is longer than $d_{\mathrm{Fe-Se}}$ by almost 0.3 \AA\ in FeSe$_{0.5}$Te$_{0.5}$. Monolayer FeSe$_{0.5}$Te$_{0.5}$ with COL-AFM order and PCB-AFM order have an interesting feature in common. $d_{\mathrm{Fe-Te}}$ in Struc.2 is shorter than those in Struc.3 and Struc.1, meanwhile, $d_{\mathrm{Fe-Se}}$ of Struc.2 is the longest among Struc.1 to Struc.3. These facts render the least difference between $d_{\mathrm{Fe-Se}}$ and $d_{\mathrm{Fe-Te}}$ in Struc.2 among the three substitutional configurations. Considering that Struc.2 is also the most stable FeSe$_{0.5}$Te$_{0.5}$ configuration seen from Tab.\ref{tab:table 2}, it seems that the longer (shorter) the $d_{\mathrm{Fe-Se}}$ ($d_{\mathrm{Fe-Te}}$), the more stable the Te-substitutional configuration. From another respect, the difference between $d_{\mathrm{Fe-Se}}$ and $d_{\mathrm{Fe-Te}}$ should be as small as possible to stabilize the substitutional structure.

\begin{table}[!bp]
\centering
\caption{\label{tab:table 3} The average distance between Se (Te) atoms and Fe-plane, denoted as $d_{\mathrm{Fe-Se}}$ ($d_{\mathrm{Fe-Te}}$), in both FeSe and FeSe$_{0.5}$Te$_{0.5}$ with different magnetic orders. Units: \AA }
\begin{ruledtabular}
\begin{tabular}{ccc|cccc}
Magnetic &\multicolumn{2}{c|}{\multirow{2}{*}{Structures}} & \multicolumn{2}{c}{No SOC} &\multicolumn{2}{c}{Including SOC} \\
Order & & & $d_{\mathrm{Fe-Se}}$ & $d_{\mathrm{Fe-Te}}$ & $d_{\mathrm{Fe-Se}}$ & $d_{\mathrm{Fe-Te}}$ \\
\hline
\multirow{2}{*}{PM} & \multicolumn{2}{c|}{FeSe}& 1.274 & & 1.274 &  \\ \cline{2-7}
&\multicolumn{2}{c|}{FeSe$_{0.5}$Te$_{0.5}$} & 1.236 & 1.544 & 1.237 & 1.546  \\  \hline
\multirow{2}{*}{CB-AFM} & \multicolumn{2}{c|}{FeSe}& 1.372 &  & 1.373 &   \\ \cline{2-7}
&\multicolumn{2}{c|}{FeSe$_{0.5}$Te$_{0.5}$} &1.363 & 1.650 & 1.365 & 1.653  \\  \hline
\multirow{4}{*}{COL-AFM} &  \multicolumn{2}{c|}{FeSe}& 1.383 & & 1.384 &  \\ \cline{2-7}
& \multirow{3}{*}{FeSe$_{0.5}$Te$_{0.5}$} & Struc.1 & 1.375 & 1.658  & 1.377 & 1.662 \\
& & Struc.2 & 1.390 & 1.651 & 1.390 & 1.658 \\
& & Struc.3 & 1.387 & 1.653 & 1.387 & 1.664 \\ \hline
\multirow{4}{*}{PCB-AFM} &  \multicolumn{2}{c|}{FeSe}& 1.395 & & 1.396 & \\ \cline{2-7}
& \multirow{3}{*}{FeSe$_{0.5}$Te$_{0.5}$} & Struc.1 & 1.387 & 1.667 & 1.387 & 1.668 \\
& & Struc.2 & 1.401 & 1.652 & 1.402 & 1.654 \\
& & Struc.3 & 1.397 & 1.663 & 1.397 & 1.665 \\
\end{tabular}
\end{ruledtabular}
\end{table}

\subsubsection{Band Structure}

\begin{figure}[!ptb]
\begin{center}
\includegraphics[width=0.5\linewidth]{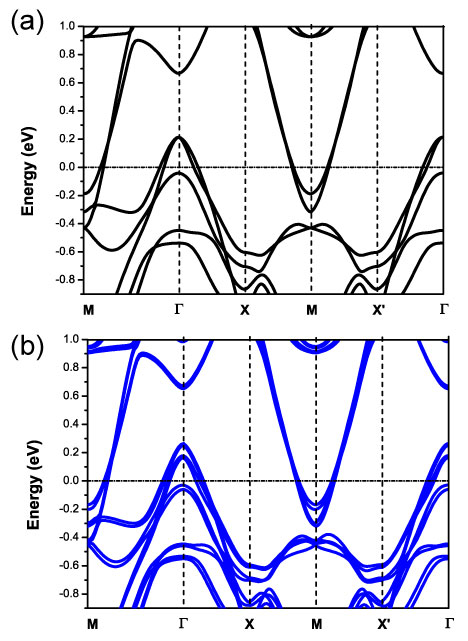}
\end{center}
\caption{(Color online) Band structures for paramagnetic FeSe$_{0.5}$Te$_{0.5}$ calculated (a) with no SOC and (b) including SOC. }
\label{Fig4}%
\end{figure}

Next, we come to the electronic band structures of FeSe$_{0.5}$Te$_{0.5}$. Figure 4 shows the band structures of paramagnetic FeSe$_{0.5}$Te$_{0.5}$, which resemble those of pure paramagnetic FeSe except that some degeneracy has been lifted. Especially at M point, the band-splitting is conspicuous, as demonstrated in Fig. 4(a). The valence bands in pure paramagnetic FeSe at M only have one maximum; in contrast, in FeSe$_{0.5}$Te$_{0.5}$ the valence bands have a maximum on each side of the M point, i.e. along M-X path and M-X' path the valence bands have symmetrical maximal points. Besides, above the valence band maximum at M, the conduction bands also have an energy-splitting in FeSe$_{0.5}$Te$_{0.5}$. The different characters of bands near the M point is undoubtedly a consequence of the Te-substitution. For a certain Fe atom, M-X direction and M-X' direction are not equivalent, since a Se atom lies on one path above Fe-plane while a Te atom lies on the other path below Fe-plane. Including SOC, the degeneracy is further lifted in the whole spectra, since SOC impacts FeSe$_{0.5}$Te$_{0.5}$ more than FeSe because a Te atom is much heavier than a Se atom, as mentioned before.

\begin{figure}[!ptb]
\begin{center}
\includegraphics[width=0.82\linewidth]{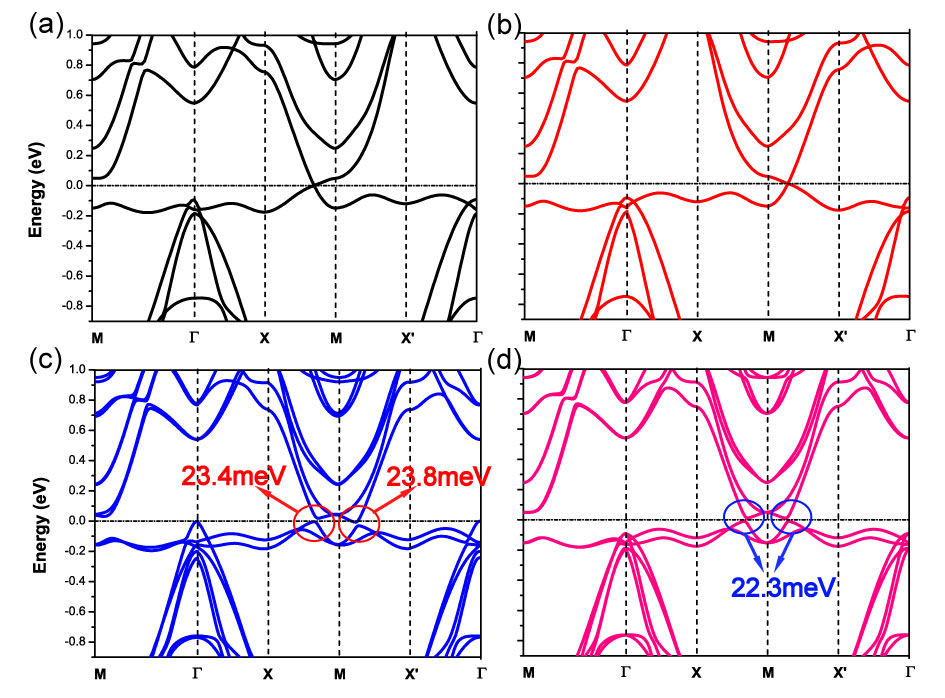}
\end{center}
\caption{(Color online) Band structures for FeSe$_{0.5}$Te$_{0.5}$ in CB-AFM order. (a) and (b) are band structures composed of electrons from different spin channels calculated without SOC. (c) and (d) are calculated including SOC with in-plane magnetization and out-of-plane magnetization, respectively.}
\label{Fig5}%
\end{figure}

Before moving to the AFM ordered FeSe$_{0.5}$Te$_{0.5}$, we recall that in pristine FeSe the two spin channels have identical band structures. But in FeSe$_{0.5}$Te$_{0.5}$ with CB-AFM order, things are different. As shown in Fig.~5(a) and (b), in each spin channel, there is a flat band below the Fermi level forming a Dirac cone which slightly deviates from M point. However, the Dirac cones are on the different sides of the M point: it is on the M-X path for one spin channel while on the M-X' path for another spin channel. These results can be explained in a similar way as comprehending the feature of the band structure of paramagnetic FeSe$_{0.5}$Te$_{0.5}$. There are two Fe ions which have opposite magnetizations in a unit cell, meaning that one Fe ion has more \emph{spin-up} electrons and the other has more \emph{spin-down} electrons. Starting from the M point, we assume that the on-top Se atoms are at the X direction of a Fe atom with more spin-up electrons. Accordingly, at the X direction of a Fe atom with more spin-down electrons lie the bottom Te atoms. As a result, for different spin channels, bands along M-X path will be different. Nevertheless, the on-top Se atoms are at the X' direction of the Fe atom with downward magnetization. Thus the M-X path for spin-up electrons should be similar to the M-X' path for spin-down electrons. But SOC diminishes the above asymmetry along X-M-X' path since the contributions from different spin channels have been added up. The flat band is split and there are gap-opening phenomena at the Dirac cones for both in-plane and out-of-plane magnetizations. Moreover, for the in-plane magnetization, the valence bands at $\Gamma$ point are pushed up to contact the Fermi level. Therefore, SOC with in-plane magnetization has altered the band structures more remarkably than SOC with out-of-plane magnetization.

The band structures of FeSe$_{0.5}$Te$_{0.5}$ with COL-AFM order are presented in Fig.~6. Compared with Fig.~3(f-h), Te-substitution has not brought any essential changes. Especially for band structures calculated without SOC, the three substitutional structures possess very similar band structures in which the Dirac cone emerges in the middle of $\Gamma$-X' path. As in pure FeSe with COL-AFM order, including SOC effect with in-plane magnetization can open a band gap at the Dirac cone, but SOC together with out-of-plane magnetization cannot open any gap in COL-AFM FeSe$_{0.5}$Te$_{0.5}$, as demonstrated in Fig.~6(d-i). However, the band gaps in FeSe$_{0.5}$Te$_{0.5}$ with COL-AFM (5.4 meV and 5.8 meV) order are smaller than the band gap in pure FeSe (8.3 meV). In fact, the SOC effect leads to various results in Struc.1 to Struc. 3. For example, the bands of Struc.2 at the Dirac cone have been merely pulled open at the Fermi level without any splitting. But a Rashba-Dresselhuas splitting is observed at the same position in band structures of Struc.1 and Struc.3. Particularly, in Struc.1 the splitting is so strong that the gap is actually closed again. As to the magnetization normal to Fe-plane, the degeneracy of bands are different in the three configurations, but the Dirac cones are all preserved.

\begin{figure*}[!ptb]
\begin{center}
\includegraphics[width=1.0\textwidth]{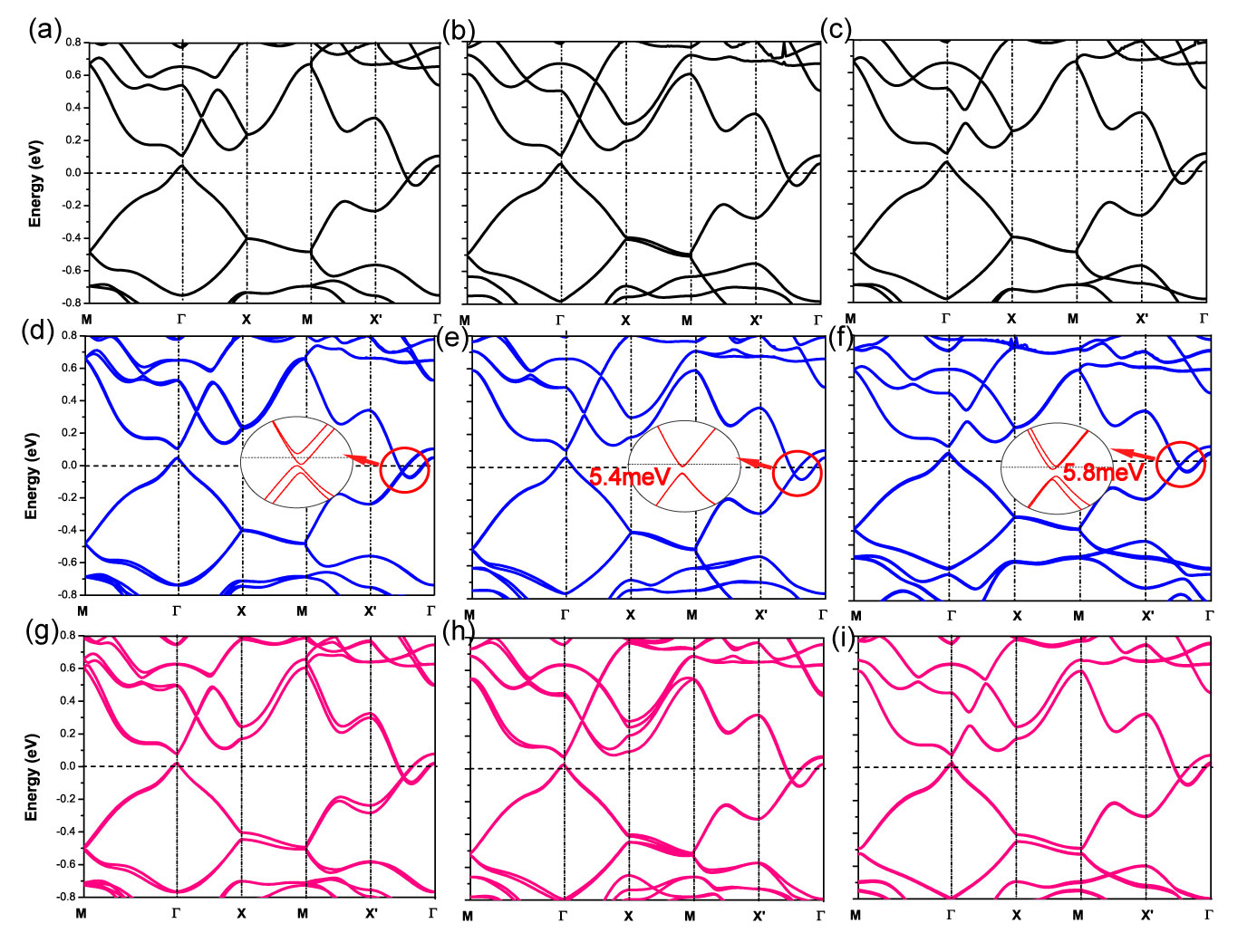}
\end{center}
\caption{(Color online) Band structures for FeSe$_{0.5}$Te$_{0.5}$ of different configurations with COL-AFM order, where (a,d,g) are band structures of Struc.1, (b,e,h) are of Struc.2 and (c,f,i) are of Struc.3. Panels (a-c) are band structures calculated without SOC. Since band structures of spin-up electrons and spin-down electrons are exactly the same, we only plot band structures for one spin channel. The bands plot in blue [panels (d-f)] and pink [panels (g-i)] are calculated including SOC with in-plane magnetization and out-of-plane magnetization, respectively. To better show the gap-opening behaviors at Dirac cones, the specific region are magnified with the gap values marked aside the gap in (d-e).}
\label{Fig6}%
\end{figure*}

\begin{figure*}[!ptb]
\begin{center}
\includegraphics[width=1.0\textwidth]{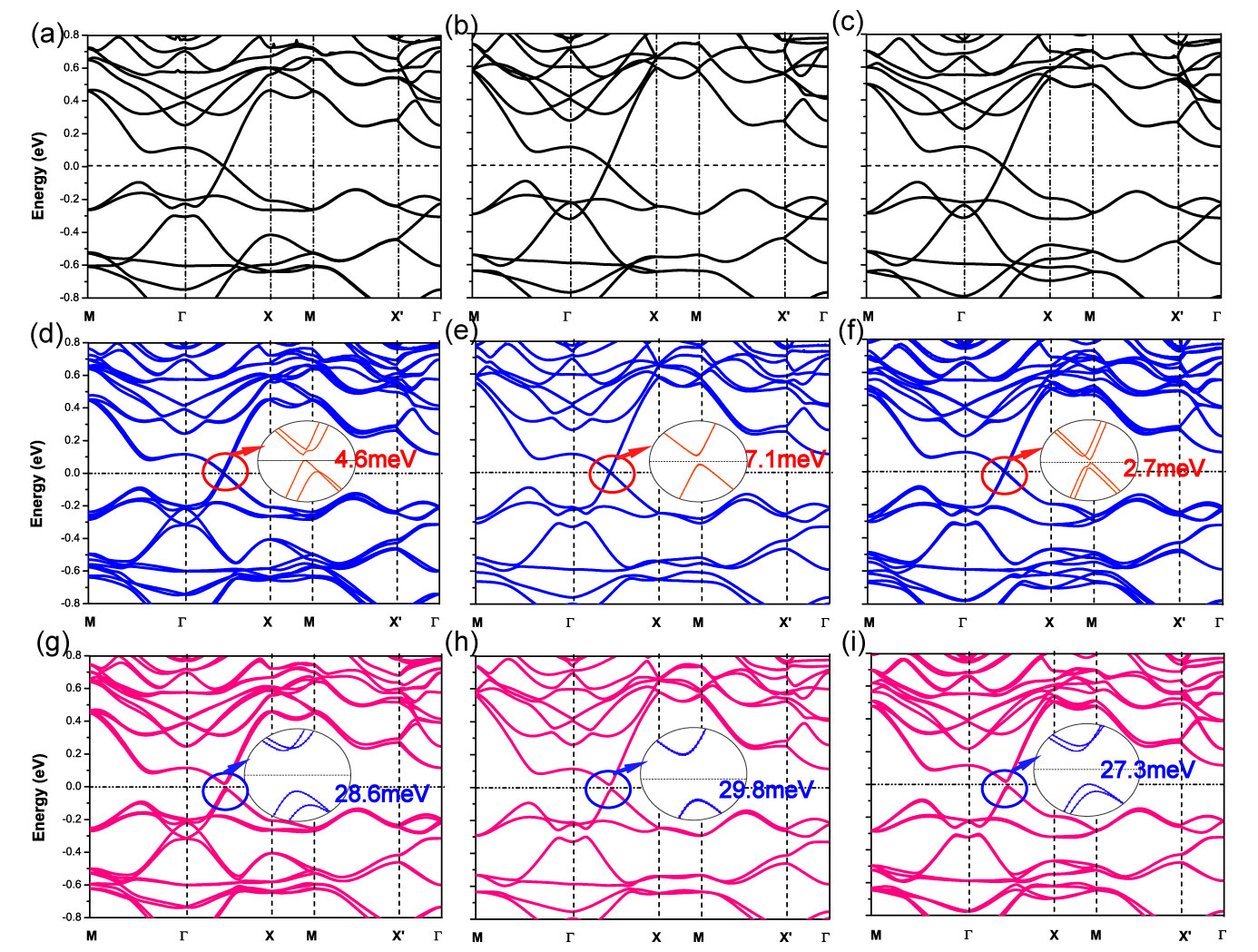}
\end{center}
\caption{(Color online) Band structures for different configurations of FeSe$_{0.5}$Te$_{0.5}$ with PCB-AFM order, where (a,d,g) are for Struc.1, (b,e,h) are for Struc.2 and (c,f,i) are for Struc.3. Panels (a-c) are band structures calculated without SOC and the bands of spin-up and spin-down electrons are totally the same. The bands plotted in blue [panels (d-f)] and pink [panels (g-i)] are calculated including SOC with in-plane magnetization and out-of-plane magnetiztion respectively. The gap-opening and gap values at Dirac cones are magnified and denoted in (d-i).}
\label{Fig7}%
\end{figure*}

Finally, the band structures of FeSe$_{0.5}$Te$_{0.5}$ with PCB-AFM order are demonstrated. Not including SOC, the band structures of FeSe$_{0.5}$Te$_{0.5}$ are similar to those of FeSe except for the valence bands at $\Gamma$ point. But these diversifications are not of major importance since they occur deep below the Fermi level. On the other hand, including SOC, the band structures manifest an insulating gap at the Dirac cone in all three substitutional structures no matter whether the magnetization is in-plane or out-of-plane, as shown in Fig. 7(d-i). When the magnetic moments are in-plane, the band gap in Struc.2 is the largest, which is 7.1 meV. In both Struc.1 and Struc.3, there are Rashba-Dresshaus splittings which lead to band gaps of 4.6 meV and 2.7 meV, respectively. If the magnetization is normal to the Fe-plane, the band gaps are larger than the gap in pure PCB-AFM FeSe by several meV. Comparing Fig. 7(d-f) with Fig. 7(g-i), one will find that SOC with out-of-plane magnetization can open band gaps which are an order of magnitude larger than the gaps opened by SOC with in-plane magnetization. However, except for the values of the band gap, the direction of the magnetization cannot influence other part of the band structures much. In summary, with the PCB-AFM order and out-of-plane magnetization, SOC effect can open sizeable band gaps in FeSe$_{0.5}$Te$_{0.5}$.

\section{Conclusion}

We have investigated the total energy, atomic structures and electronic band structures of monolayer FeSe$_{0.5}$Te$_{0.5}$ as well as monolayer FeSe with paramagnetic state and different antiferromagnetic orders using first-principles calculations. The collinear (COL) AFM order is found to be the magnetic ground state for pure FeSe monolayer. For the Te-substituted configurations, i.e. FeSe$_{0.5}$Te$_{0.5}$, the pair-checkerboard (PCB) AFM order is always the most stable magnetic state. The band structures of AFM ordered FeSe and FeSe$_{0.5}$Te$_{0.5}$ all have Dirac-cone-like features. Including spin-orbit coupling, the Dirac cones could be opened or could not, determined by the direction of the magnetic moments (in-plane or out-of-plane). It is clear that SOC effect is stronger in FeSe$_{0.5}$Te$_{0.5}$ since Te atoms are heavier than Se atoms. Besides, because of the symmetry breaking brought by Te-substitution, the band structures have new characters of Rashba-Dresselhaus splitting in FeSe$_{0.5}$Te$_{0.5}$ with COL-AFM and PCB-AFM orders.
But the band gaps of FeSe$_{0.5}$Te$_{0.5}$ are not enlarged much compared to FeSe, which is in accordance with experiments where the superconductor gap and $T_{C}$ are not enhanced in FeSeTe.

\section*{ACKNOWLEDGEMENTS}
This work was supported by NSFC under Grant No. 11474030, and by special program for applied research on super computation of the NSFC-TJ National Supercomputing Center. Li greatly appreciates the financial support of General Financial Grant from the China Postdoctoral Science Foundation (Grant Nos. 2016M591123) and also acknowledge generous grants of the Explorer 100 cluster system of Tsinghua University.


\begin{thebibliography}{27}%


\bibitem{monolayer1} L. Z. Deng, B. Lv, Z. Wu, Y. Y. Xue, W. H. Zhang, F. S. Li, L. L. Wang, X. C. Ma, Q. K. Xue, and C. W. Chu,
\newblock Phys. Rev. B \textbf{90}, 214513 (2014).


\bibitem{monolayer2} R. Peng, H. C. Xu, S. Y. Tan, H. Y. Cao, M. Xia, X. P. Shen, Z. C. Huang, C. H. P. Wen, Q. Song, T. Zhang, B. P. Xie, X. G. Gong, and D. L. Feng,
\newblock Nat. Commun. \textbf{5}, 5044 (2014).

\bibitem{monolayer3} J.-F. Ge, Z.-L. Liu, C. Liu, C.-L. Gao, D. Qian, Q.-K. Xue, Y. Liu, and J.-F. Jia,
\newblock Nat. Mater. \textbf{14}, 285 (2015).

\bibitem{berlijn2014} T. Berlijn, H.-P. Cheng, P. J. Hirschfeld, and W. Ku,
\newblock Phys. Rev. B \textbf{89}, 020501 (2014).

\bibitem{Cao2014} H.-Y. Cao, S. Tan, H. Xiang, D. L. Feng, and X.-G. Gong,
\newblock Phys. Rev. B \textbf{89}, 014501 (2014).

\bibitem{Cui2015} Y.-T. Cui, R. G. Moore, A.-M. Zhang, Y. Tian, J. J. Lee, F. T. Schmitt, W.-H. Zhang, W. Li, M. Yi, Z.-K. Liu, M. Hashimoto, Y. Zhang, D.-H. Lu,
 T. P. Devereaux, L.-L. Wang, X.-C. Ma, Q.-M. Zhang, Q.-K. Xue, D.-H. Lee, and Z.-X. Shen
\newblock Phys. Rev. Lett. \textbf{114}, 037002 (2015).

\bibitem{interfaceARPES} J. J. Lee,	F. T. Schmitt, R. G. Moore, S. Johnston, Y.-T. Cui,	W. Li, M. Yi, Z. K. Liu, M. Hashimoto, Y. Zhang, D. H. Lu, T. P. Devereaux, D.-H. Lee, and Z.-X. Shen,
\newblock Nature \textbf{515}, 245 (2014).


\bibitem{zheng} F. Zheng, Z. Wang, W. Kang, P. Zhang
\newblock Sci. Rep. \textbf{3}, 2213 (2013).

\bibitem{xiangtao} K. Liu, Z.-Y. Lu, and T. Xiang
\newblock Phys. Rev. B \textbf{85}, 235123 (2012).

\bibitem{Cao2015} H.-Y. Cao, S. Chen, H. Xiang, and X.-G. Gong,
\newblock Phys. Rev. B \textbf{91}, 020504 (2015).

\bibitem{Lu} K. Liu, Z.-Y. Lu, and T. Xiang,
\newblock Phys. Rev. B \textbf{93}, 205154 (2015).

\bibitem{neutron} F. Ye, S. Chi, Wei Bao, X. F. Wang, J. J. Ying, X. H. Chen, H. D. Wang, C. H. Dong, and Minghu Fang,
\newblock Phys. Rev. Lett. \textbf{107}, 137003 (2011).

\bibitem{squid} Y. Mizuguchi, F. Tomioka1, S. Tsuda1, T. Yamaguchi,
 and Y. Takano,
\newblock Appl. Phys. Lett. \textbf{93}, 152505 (2008).

\bibitem{ARPES2} S. Tan, Y. Zhang, M. Xia, Z. Ye, F. Chen, X. Xie, R. Peng, D. Xu, Q. Fan, H. Xu, J. Jiang, T. Zhang, X. Lai,
T. Xiang, J. Hu, B. Xie, and D. Feng
\newblock Nat. Mater. \textbf{12}, 634 (2013).


\bibitem{FengLiu} Z.F. Wang, H. Zhang, D. Liu, C. Liu, C. Tang, C. Song, Y. Zhong, J. Peng, F. Li, C. Nie, L. Wang,	X.J. Zhou, X. Ma, Q.K. Xue, and F. Liu,
\newblock Nat. Mater. \textbf{15}, 968 (2016).

\bibitem{cohen} T. Bazhirov, and M. L. Cohen,
\newblock J. Phys. Condens. Mat. \textbf{25}, 105506 (2013).

\bibitem{zheng2} F. Zheng, L. Wang, Q.-K. Xue, and P. Zhang,
\newblock Phys. Rev. B \textbf{93}, 075428 (2016).

\bibitem{bulkTC} F. C. Hsu, J. Y. Luo, K. W. The, T. K. Chen, T. W. Huang, P. M. Wu, Y. C. Lee, Y. L. Huang, Y. Y. Chu, D. C. Yan, and M. K. Wu,
\newblock Proc. Natl. Acad. Sci. USA \textbf{105}, 14262 (2008).

\bibitem{FeSeTeTC} M. H. Fang, H. M. Pham, B. Qian, T. J. Liu, E. K. Vehstedt, Y. Liu, L. Spinu, and Z. Q. Mao,
\newblock Phys. Rev. B \textbf{78}, 224503 (2008).

\bibitem{FeSeTeTC2} T. Hanaguri, S. Niitaka, K. Kuroki, and H. Takagi,
\newblock Science \textbf{328}, 474 (2010).



\bibitem{external} S. Kawale, E. Bellingeri, V. Braccini, I. Pallecchi, M. Putti, G. Grimaldi,
A. Leo, A. Guarino, A. Nigro, and C. Ferdeghini,
\newblock IEEE Trans. Appl. Supercond. \textbf{23}, 7500704 (2013).

\bibitem{topoFeSeTe} Y. Xia, D. Qian, L. Wray, D. Hsieh, G. F. Chen, J. L. Luo, N. L. Wang, and M. Z. Hasan,
\newblock Phys. Rev. Lett. \textbf{103}, 037002 (2009).

\bibitem{paw1} P. E. Blochl,
\newblock Phys. Rev. B {\bf 50}, 17953 (1994).

\bibitem{paw2}  G. Kresse, and D. Joubert,
\newblock Phys. Rev. B {\bf 59}, 1758 (1999).

\bibitem{vasp}  G. Kresse, and J. Furthm$\ddot{u}$ller,
\newblock Phys. Rev. B {\bf 54}, 11169 (1996).

\bibitem{pbe1} J.~P. Perdew, K. Burke, and M. Ernzerhof,
\newblock Phys. Rev. Lett. {\bf 77}, 3865 (1996).

\bibitem{cg} W.H. Press, B.P. Flannery, S.A. Teukolsky and W.T. Vetterling,
\newblock em Numerical Recipes (Cambridge University Press, New York, 1986).

%
%
%
%
%
%
%

%
%



\end{thebibliography}
\end{document}